# Thermal Properties of Graphene, Carbon Nanotubes and Nanostructured Carbon Materials

## Alexander A. Balandin

Department of Electrical Engineering and Materials Science and Engineering Program,
Bourns College of Engineering, University of California, Riverside, CA 92521 U.S.A.

E-mail: balandin@ee.ucr.edu
Web: http://ndl.ee.ucr.edu

### Abstract

Recent years witnessed a rapid growth of interest of scientific and engineering communities to thermal properties of materials. Carbon allotropes and derivatives occupy a unique place in terms of their ability to conduct heat. The room-temperature thermal conductivity of carbon materials span an extraordinary large range – of over five orders of magnitude – from the lowest in amorphous carbons to the highest in graphene and carbon nanotubes. I review thermal and thermoelectric properties of carbon materials focusing on recent results for graphene, carbon nanotubes and nanostructured carbon materials with different degrees of disorder. A special attention is given to the unusual size dependence of heat conduction in two-dimensional crystals and, specifically, in graphene. I also describe prospects of applications of graphene and carbon materials for thermal management of electronics.

### I. Introduction

Recently increasing importance of thermal properties of materials is explained both by practical needs and fundamental science. Heat removal has become a crucial issue for continuing progress in electronic industry owing to increased levels of dissipated power. The search for materials that conducts heat well became essential for design of the next generations of integrated circuits (ICs) and three-dimensional (3D) electronics [1]. Similar thermal issues have been encountered in optoelectronic and photonic devices. From another side, thermoelectric energy conversion requires materials, which have strongly suppressed thermal conductivity $K$ [2].

The material's ability to conduct heat is rooted in its atomic structure, and knowledge of thermal properties can shed light on other materials' characteristics. Thermal properties of materials change when they are structured on a nanometer scale. Nanowires do not conduct heat as well as bulk crystals due to increased phonon - boundary scattering [3] or changes in the phonon dispersion [4]. At the same time, theoretical studies of heat conduction in two-dimensional (2D) and one-dimensional (1D) crystals revealed exotic behavior, which lead to infinitely large intrinsic thermal conductivity [5-6]. The $K$ divergence in 2D crystals means that unlike in bulk, the crystal anharmonicity alone is not sufficient for restoring thermal equilibrium, and one needs to either limit the system size or introduce disorder to have the physically meaningful *finite* value of $K$. These findings led to discussions of the validity of Fourier's law in low-dimensional systems [7-8].

Carbon materials, which form a variety of allotropes [9], occupy a unique place in terms of their thermal properties (see Figure 1a). Thermal conductivity of different allotropes of carbon span an extraordinary large range – of over five orders of magnitude – from the lowest of ~ 0.01 W/mK in amorphous carbon (*a*-C) to the highest of above 2000 W/mK at room temperature (RT) in diamond or graphene. In type-II-*a* diamond $K$ reaches 10000 W/mK at $T \approx 77$ K. Thermal conductivity of carbon nanotubes (CNTs), $K \approx 3000 - 3500$ W/mK at RT [10-11], exceeds that of diamond, which is the best bulk heat conductor.

The exfoliation of graphene [12] and discovery of its exotic electrical conduction [13-15] made possible, among other things, the first experimental study of heat transport in the strictly 2D crystals. Availability of high-quality few-layer graphene (FLG) led to experimental observations of evolution of thermal properties as the system dimensionality changes from 2D to 3D. The first measurements of thermal properties of graphene [16-19], which revealed $K$ above the bulk graphite limit, ignited strong interest to thermal properties of this material and, in a more general context, to heat conduction in crystals of lower dimensionality. Rapidly increasing number of publications on the subject, often with contradictory results, calls for a comprehensive review. Such a review with the emphasis on graphene is particularly appropriate because this material provided the recent stimulus for thermal research, and it likely holds the keys to understanding heat conduction in low dimensions. These considerations motivated the present work, which discusses thermal properties of graphene and CNTs in the context of carbon allotropes.

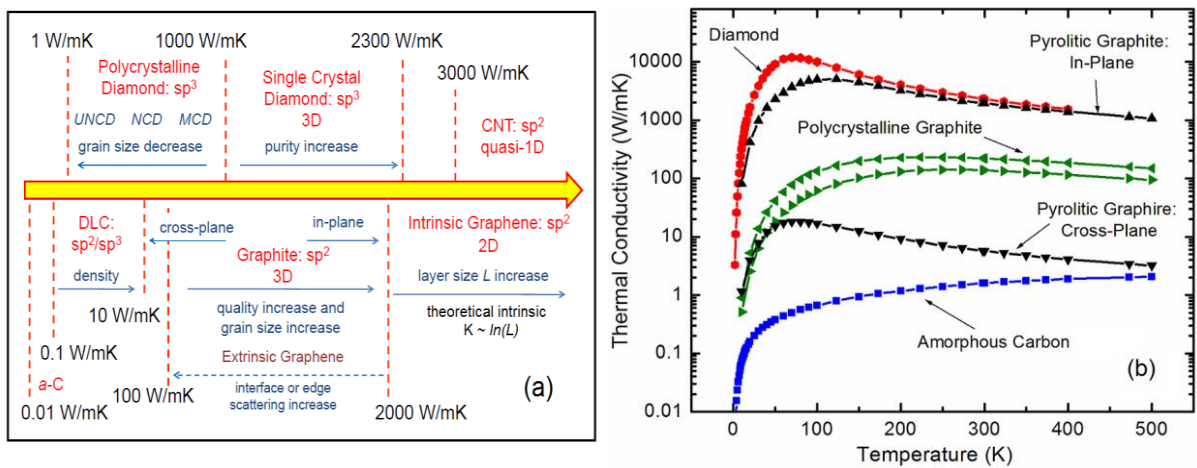

**Figure 1**: Thermal properties of carbon allotropes and derivatives. (a) Diagram based on average values reported in literature. The axis is not to scale. (b) Thermal conductivity of bulk carbon allotropes as a function of $T$. The plots are based on commonly accepted *recommended* values from ref. (29). The curve "Diamond" is for the electrically insulating diamond of type II-$a$; "Polycrystalline Graphite" is for AGOT graphite – a high-purity pitch-bonded graphite; and "Pyrolytic Graphite" is for high-quality graphite analog to HOPG. Note an *order of magnitude* difference in $K$ of pyrolytic graphite and polycrystalline graphite with disoriented grains. The $K$ value for pyrolytic graphite constitutes the bulk graphite limit of ~2000 W/mK at RT. At low $T$, $K$ is proportional to $T^\gamma$, where $\gamma$ varies in a wide range depending on graphite's quality and crystallites size [29-30].

## II. Basics of Heat Conduction: Bulk vs. Nanostructures

Here, I define main quantities and outline the nanoscale size effects on heat conduction. Thermal conductivity is introduced through Fourier's law, $\vec{q} = -K\nabla T$, where $\vec{q}$ is the heat flux and $\nabla T$ is the temperature gradient. In this expression, $K$ is treated as a constant, which is valid for small $T$ variations. In a wide temperature range, $K$ is a function of $T$. In anisotropic materials, $K$ varies with crystal orientation and is represented by a tensor [20-22].

In solid materials heat is carried by acoustic phonons, i.e. ion core vibrations in crystal lattice, and electrons so that $K=K_p+K_e$, where $K_p$ and $K_e$ are the phonon and electron contributions, respectively. In metals, $K_e$ is dominant due to large concentration of free carriers. In pure copper − one of the best metallic heat conductors − $K\approx400$ W/mK at RT and $K_p$ is limited to 1-2% of the total. Measurements of the electrical conductivity $\sigma$ define $K_e$ via the Wiedemann-Franz law, $K_e/(\sigma T) = \pi^2 k_B^2/(3e^2)$, where $k_B$ is Boltzmann constant and $e$ is the charge of an electron. Heat conduction in carbon materials is usually dominated by phonons even for graphite [23], which has metal-like properties [24]. The latter is explained by the strong covalent $sp^2$ bonding resulting in efficient heat transfer by lattice vibrations. However, $K_e$ can become significant in doped materials.

The phonon thermal conductivity is expressed as $K_p = \Sigma_j \int C_j(\omega)\upsilon_j^2(\omega)\tau_j(\omega)d\omega$, where $j$ is the phonon polarization branch, i.e. two transverse acoustic (TA) and one longitudinal acoustic (LA) braches, $\upsilon$ is the phonon group velocity, which, in many solids, can be approximate by the sound velocity, $\tau_j$ is the phonon relaxation time and $C_j$ is the contribution to heat capacity from the given branch $j$. The phonon mean-free path (MFP) $\Lambda$ is related to the relaxation time as $\Lambda = \tau\upsilon$. In the relaxation-time approximation (RTA), various scattering mechanisms, which limit MFP, are additive, i.e. $\tau^{-1} = \Sigma\tau_i^{-1}$. In typical solids, the acoustic phonons, which carry bulk of heat, are scattered by other phonons, lattice defects, impurities, conduction electrons and interfaces [22, 25-26]. A simpler equation for $K$, derived from the kinetic theory of gases, is $K_p = (1/3)C_p\upsilon\,\Lambda$.

It is important to distinguish between *diffusive* and *ballistic* phonon transport regimes. The thermal transport is called *diffusive* if the size of the sample $L$ is much larger than $\Lambda$, i.e. phonons undergo many scattering events. When $L < \Lambda$ the thermal transport is termed *ballistic*. Fourier law assumes diffusive transport. Thermal conductivity is called *intrinsic* when it is limited by the crystal lattice anharmonicity. The crystal lattice is anharmonic when its potential energy has terms higher than the second order with respect to the ion displacements from equilibrium. The intrinsic $K$ limit is reached when the crystal is perfect, without defects or impurities, and phonons can only be scattered by other phonons, which "see" each other due to anharmonicity. The anharmonic phonon interactions, which lead to finite $K$ in 3D, can be described by the Umklapp processes [22]. The degree of crystal anharmonicity is characterized by the Gruneisen parameter $\gamma$, which enters the expressions for the Umklapp scattering rates [22, 25]. Thermal conductivity is *extrinsic* when it is mostly limited by the extrinsic effects such phonon – rough boundary or phonon – defect scattering.

In nanostructures $K$ is reduced by scattering from boundaries, which can be evaluated as [27] $1/\tau_B = (\upsilon / D)[(1-p)/(1+p)]$. Here $D$ is the nanostructure or grain size and $p$ is the *specularity* parameter defined as a probability of specular scattering at the boundary. The momentum-conserving specular scattering ($p$=1) does not add to thermal resistance. Only diffuse phonon scattering from rough interfaces ($p\rightarrow0$), which changes the momentum, limits the phonon MFP. One can find $p$ from the surface roughness or use it as a fitting parameter to experimental data. When the phonon - boundary scattering is dominant, $K$ scales with $D$ as $K_p \sim C_p \upsilon \Lambda \sim C_p \upsilon^2 \tau_B \sim C_p \upsilon D$. In nanostructures with $D<<\Lambda$, phonon dispersion can undergo modifications due to confinement resulting in changes in $\upsilon$ and more complicated size dependence [28]. The specific heat $C_p$ is defined by the phonon density of states (DOS), which leads to different $C_p(T)$ dependence in 3D, 2D and 1D systems, and reflected in $K(T)$ dependence at low $T$ [22, 27]. For example, in bulk at low $T$, $K(T) \sim T^3$ while it is $K(T)\sim T^2$ in 2D systems.

### III. Thermal Properties of Bulk Carbon Allotropes

I start by revisiting thermal properties of bulk carbon allotropes – graphite, diamond and amorphous carbon. This provides proper reference for the discussion of graphene and nanostructured carbons. It also helps to distinguish new physics emerging in low-dimensional structures from mundane material quality issues. It is hard to find another material, where $K$ was studied as rigorously as in graphite. One of the reasons for it was the needs of nuclear industry. Ironically, the data for graphite sometimes are difficult to find because the studies were conducted in last century and often published in reports of limited circulation. Correspondingly, there is confusion among modern-days researchers what is $K$ value for basal planes of high-quality graphite. Figure 1b shows thermal conductivity of two types of high purity graphite (sp$^2$ bonding), diamond (sp$^3$) and amorphous carbon (disordered mixture of sp$^2$/sp$^3$). These plots are based on the *recommended* values reported in ref. (29), which were obtained by compilation and analysis of *hundreds* of research papers with conventionally accepted experimental data.

The pyrolytic graphite, which is similar to the highly-oriented pyrolytic graphite (HOPG), has in-plane $K{\approx}2000$ W/mK at RT. Its cross-plane $K$ is more than two orders of magnitude smaller. Another type of chemically-pure pitch-bonded graphite, produced by different technique, has an *order of magnitude* smaller in-plane $K{\sim}200$ W/mK. The $K$ anisotropy is much less pronounced. This difference is explained by the fact that HOPG is made from large crystallites, which are well aligned with each other, so that the overall behavior is similar to that of a single crystal [30]. Pitch-bonded graphite is also polycrystalline, but the crystal axes are not highly oriented and the grain boundaries are more pronounced [30]. As a result, $K$ of polycrystalline graphite of the types other than HOPG can be strongly limited by the grain size $D$. The same factors limit $K$ in the chemical vapor deposited (CVD) graphene, which is polycrystalline with misoriented grains [31-32]. Thus, I consider $K{\sim}2000$ W/mK as the graphite RT bulk limit. Any smaller value is indicative of the lower quality graphite, where $K$ is limited by phonon scattering on grain boundaries, defects or rough sample edges. The experimental $K$ values for HOPG are in excellent agreement with theoretical predictions for the intrinsic $K$ of graphite [23, 33].

In all bulk carbon allotropes, heat is mostly carried by acoustic phonons. In diamond and HOPG, $K$ attains its maximum at ~70 K and ~100 K, respectively. For higher $T$, $K$ decreases as ~$1/T$, which is characteristic of crystalline solids, where $K$ is limited by phonon Umklapp scattering. Thermal conductivity of *a-C* varies from ~0.01 W/mK at $T$=4K to ~2 W/mK at $T$=500 K. It increases monotonically with $T$, which is expected for disordered materials, where the heat conduction mechanism is hopping of localized excitations [34]. As seen from Figure 1b, $K$ for HOPG and pitch-bonded graphite has different $T$ dependence at low temperature. It is well known that $K(T)$ dependence in graphite varies substantially. It is defined not only by the phonon DOS via $C_p$ but also by graphite's grain size and quality [29-30].

## IV. Disordered and Nanostructured Carbon Materials

Let us discuss thermal properties of materials where $K$ is limited by disorder or grain boundaries rather than by the intrinsic lattice dynamics. An important representative of this class of materials is diamond-like carbon (DLC), which is a metastable form of $a$-$C$ containing a significant fraction of sp$^3$ bonds [35]. DLC films are widely used as protective coatings with optical windows for magnetic storage disks and in biomedical applications. DLC consists not only of $a$-$C$ but also of the hydrogenated alloys, $a$-$C$:$H$. Hydrogen-free DLC with the highest sp$^3$ content are called tetrahedral amorphous carbon ($ta$-$C$). Experimental studies [36-42] revealed that heat conduction in DLC is mostly governed by the amount and structural disorder of sp$^3$ phase. If sp$^3$ phase is amorphous, $K$ scales approximately linearly with the sp$^3$ content, density, and elastic constants (Figure 2a). The polymeric and graphitic DLC films have the lowest $K \sim 0.1$–$0.3$ W/mK, hydrogenated $ta$-$C$:$H$ has $K \sim 1$ W/mK, and $ta$-$C$ has the highest $K$, which can go up to $\sim 10$ W/mK at RT [41]. Among amorphous solids, $ta$-$C$ is likely the material with the highest $K$ [35-36]. If the sp$^3$ phase orders, even in small grains such as in nanocrystalline diamond, a strong $K$ increase occurs for a given density $\rho$, Young's modulus $E$, and sp$^3$ content.

Progress in CVD polycrystalline diamond ($p$-D) films – ultrananocrystalline (UNCD), nanocrystalline (NCD) and microcrystalline (MCD) (Figure 2b) – renewed interest to their thermal properties [43-44]. Most studies of $p$-D [45-51] agree that $K$ depends strongly on the grain size $D$ and cover the range from $\sim 1$-$10$ W/mK in UNCD to $\sim 550$ W/mK in MCD ($D \sim 3$-$4\ \mu$m). The grain size $D$ dependence can be estimated from $K_p \approx (1/3) C \upsilon D$, which assumes that inside the grain, phonon propagation is the same as in bulk crystal. The latter was confirmed by the high resolution measurements of the local thermal conductivity in $p$-D [46-47]. More accurate theoretical description can be achieved with RTA via introduction of the scattering on grain boundaries and defects inside the grains [52]. The phonon-hopping model [53], which involves phonon transmission through grain boundaries, gave good agreement for *p*-D with different *D* (Figure 2c). Some studies suggested that heat conduction can be different in UNCD with ultra-small *D*~3-5 nm, where thermal transport is controlled by the intrinsic properties of the grain boundaries [49]. The grain boundaries contain $sp^2$-phase as opposed to the $sp^3$-phase carbon inside the grains [54].

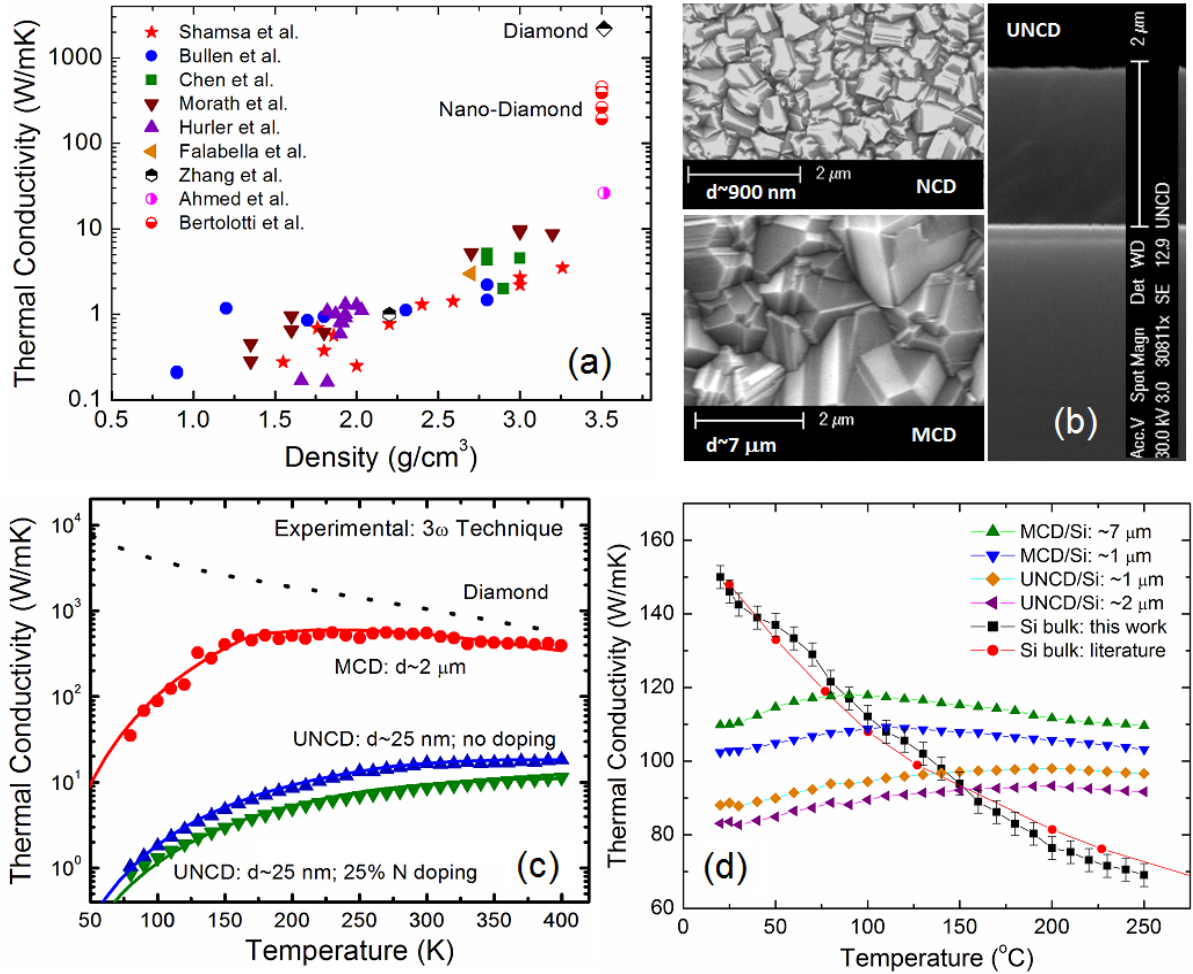

**Figure 2:** Thermal conductivity of disordered and nanostructured carbon materials. (a) *K* of DLC as a function of mass density. Note that ordering of $sp^3$ phase inside grains in NCD results in significant *K* increase. (b) Scanning electron microscopy images showing the UNCD - Si interface and grain sizes in NCD and MCD. (c) Comparison of *K* temperature dependence for UNCD and MCD films. (d) Effective thermal conductivity of MCD/Si and UNCD/Si composite substrates indicating that they can outperform Si wafers at elevated *T* in terms of thermal properties. Figures (a), (c) and (d) are adapted from refs. (41), (42) and (55), respectively.

Applications of *p*-D films for heat spreading in ICs can become feasible if *K* of the composite Si/*p*-D substrate becomes smaller than that of Si wafers. There are trade-offs in optimizing Si/*p*-D substrates. MCD films have higher *K* because of the larger grains but suffer from rough interface with Si, which increases thermal resistance of the structure. UNCD form better interfaces but have few-nm size grains. Recent developments mark progress in this direction (Figure 2d). It has been shown that composite Si/*p*-D substrates, which are more thermally resistive at RT, outperform Si wafers at elevated temperatures (above *T*~360 K), which are characteristic for operation of electronic devices [55].

## V. Thermal Conductivity of Carbon Nanotubes

Thermal transport in CNTs and graphene, unlike in NCD or DLC, can be dominated by the intrinsic properties of the strong sp$^2$ lattice, rather than by phonon scattering on boundaries or by disorder, giving rise to extremely high *K* values [10,11, 16, 17]. From the theoretical point of view, CNTs are similar to graphene but have large curvatures and different quantization conditions for phonon modes. In the discussion of heat conduction in CNTs, one has to take into account the ambiguity of the intrinsic *K* definition for 2D and 1D systems [5-8, 56-62]. Although graphene is structurally simpler, I start with experimental data for CNTs, because their thermal properties have been studied for more than a decade. CNTs became the first nanostructures with reported experimental *K* exceeding that of bulk graphite and diamond.

Table I summarizes experimental data for single-wall (SW) and multi-wall (MW) CNTs [10-11, 63-66]. Theoretical results [67-70] are also provided for comparison. There is substantial data scatter in the reported RT values for CNTs ranging from *K*≈1100 W/mK [71] to *K*≈7000 W/mK [64]. The highest *K* values obtained in the experiments were attributed to the *ballistic* transport regime achieved in some CNTs. Commonly quoted values for individual MW-CNTs are ~3000 W/mK [10] and ~3500 W/mK for SW-CNTs [11] at RT. These values are above the bulk graphite limit of ~2000 W/mK. Thus, CNTs are nanostructures where heat transport is not mostly limited by the extrinsic effects, such as boundary scattering, like in many semiconductor nanowires with rough interfaces.

The largest phonon MFP extracted from the measurements [10, 64] was *Λ*~700 − 750 nm at RT. Since the length of the measured CNTs *L* was above 2 μm, the phonon transport was still diffusive but close to the ballistic transition. At $T<30$ K the energy-independent phonon MFP of $\sim 0.5 - 1.5$ mm was extracted from measurements for SW-CNT bundles [63]. The peak in CNT $K$ was achieved at $T\sim 320$ K [10], which is substantially higher $T$ compared to bulk crystals. This indicates that Umklapp phonon scattering is suppressed in CNTs over a wide $T$ range. At low $T$, $K(T)$ follows the temperature dependence of $C_p$. For individual MW-CNTs, $K(T) \sim T^{2.5}$ was observed [10], which is similar to bulk graphite [29]. In SW-CNT bundles the $K(T)$ dependence was linear for $T<30$ K [63]. The thermoelectric measurements with SW-CNTs revealed Seebeck coefficient $S\approx 42$ μV/K at RT, which is about an order of magnitude higher than that of graphite or metals suggesting that electron transport is not ballistic [64].

**Table I: Thermal Conductivity of Graphene and Carbon Nanotubes**

| Sample | K (W/mK) | Method | Comments | Refs |
|--------|----------|--------|----------|------|
| MW-CNT | >3000 | electrical; micro-heater | individual; diffusive; suspended | 10 |
| SW-CNT | ~3500 | electrical self-heating | individual; boundary | 11 |
| SW-CNTs | 1750 – 5800 | thermocouples | bundles; diffusive | 63 |
| SW-CNT | 3000 – 7000 | electrical; micro-heater | individual; ballistic; suspended | 64 |
| CNT | 1100 | electrical; micro-heater | individual; suspended | 71 |
| CNT | 1500 - 2900 | electrical | individual | 65 |
| CNT | ~6600 | Theory: MD | $K_{CNT} < K_{graphene}$ | 66 |
| CNT | ~3000 | Theory: MD | strong defect dependence | 67 |
| SW-CNT | ~2500 | Theory: BTE | $K_{CNT} < K_{graphene}$ | 69 |
| SW-CNT | ~7000 | Theory: MD and BTE | $L$>20 nm | 70 |
| graphene | ~2000 - 5000 | Raman optothermal | suspended; exfoliated | 16 |
| FLG | ~1300 - 2800 | Raman optothermal | suspended; exfoliated; $n$=4-2 | 74 |
| graphene | ~2500 | Raman optothermal | suspended; CVD | 75 |
| graphene | 1500 - 5000 | Raman optothermal | suspended; CVD | 77 |
| graphene | 600 | Raman optothermal | suspended; exfoliated; $T$~660 K | 76 |
| FLG ribbon | 1100 | electrical self-heating | supported; exfoliated; $n$<5 | 79 |
| graphene | 600 | electrical | supported; exfoliated | 78 |
| graphene | 2000 - 5000 | Theory: VFF, BTE, γ(q) | strong width dependence | 83 |
| graphene | 1000 - 5000 | Theory: RTA, $\gamma_{TA}$, $\gamma_{LA}$ | strong size dependence | 62 |
| graphene | 8000 - 10000 | Theory: MD, Tersoff | square graphene sheet | 84 |
| graphene | 1400 - 2400 | Theory: BTE | length dependence | 69 |
| graphene | ~4000 | Theory: ballistic | strong width dependence | 86 |

[*]VFF – valence force field, BTE – Boltzmann transport equation, RTA – relaxation time approximation, MD – molecular dynamics, γ(q) – Gruneisen parameter dependent on the phonon wave vector, $\gamma_{LA}$ and $\gamma_{TA}$ – Gruneisen parameter averaged separately for LA and TA phonon modes. The data is for near RT unless otherwise specified in the Comments column.

The measured thermal conductance $G_p$ in SW-CNT was found to increase with $T$ from $0.7\times 10^{-9}$ W/K or $\sim 7g_0$ at 110K to $3.8\times 10^{-9}$ W/K or $\sim 14g_0$ at RT [64], where $g_0=\pi^2 k_B^2 T/3h\approx (9.456\times 10^{-9}$ W/K$^2)T$ is the universal quantum of thermal conductance and represent the maximum possible conductance per phonon mode [72]. Assuming different CNT diameter $d_{CNT}$ in the range from 1 nm to 3 nm, the extracted $K$ of SW-CNTs was found to change from ~8000 to ~2500 W/mK at RT [64]. An experimental study reported decreasing $K$ in MW-CNTs from ~2800 to ~500 W/mK with the outer diameter increasing from 10 nm to ~28 nm [65]. The same $K$ dependence on CNT diameter was reported in ref. (71). This experimental trend for MW-CNTs suggests that the interactions of phonons and electrons between multi-walled layers affect $K$. Thermal conductivity increases as the number of atomic walls in MW-CNTs decreases [65]. Interestingly, the Boltzmann transport equation (BTE) predicts increasing $K$ with increasing diameter for SW-CNTs when $1 < d_{CNT} < 8$ nm [69].

### VI. Experimental Studies of Graphene

The first experimental studies [16-17, 73-74] of thermal conductivity of graphene were carried out at UC-Riverside (see details of the measurement procedures in Appendix I). The optothermal Raman measurements were performed with large-area suspended graphene layers exfoliated from high-quality HOPG. The authors found $K$ exceeding ~3000 W/mK near RT, i.e. above the bulk graphite limit, observed $K$ dependence on the layer size and determined that $K_e << K_p$. The phonon MFP was estimated to be $\Lambda \sim 775$ nm near RT [17]. A following independent study [75] also utilized the Raman technique but modified it via addition of a power meter under the suspended portion of graphene. It was found that $K$ of suspended high-quality CVD graphene exceeded ~2500 W/mK at 350 K, and it was as high as $K \approx 1400$ W/mK at 500 K (experimental uncertainty ~40%) [75]. The reported value is larger than $K$ of bulk graphite at RT. Another group that repeated the optothermal Raman measurements found $K \approx 630$ W/mK for suspended graphene at $T \approx 600$ K [76]. The graphene membrane was heated to $T=660$ K in the center and above ~500 K over most of its area. Since $K$ decreases with $T$, this fact can explain the difference with refs. (16-17, 75), which reported $K$ near RT. Differences in strain distribution in the suspended graphene of various sizes and geometries may also affect the results. Other optothermal studies with suspended graphene found $K$ in the range from ~1500 to ~5000 W/mK [77].

The data for suspended or partially suspended graphene is closer to the intrinsic $K$ because suspension reduces thermal coupling to the substrate and scattering on the substrate defects and impurities. It also helps to form the in-plane heat wave front, which allows one to obtain the data pertinent to graphene itself rather than to the graphene-substrate interface even if only a part of the layer is suspended. For practical applications, it is important to know $K$ of supported graphene, i.e. graphene attached to the substrate along its entire length. The measurements for exfoliated graphene on $SiO_2$/Si revealed in-plane $K\approx600$ W/mK near RT [78]. This value is below those reported for suspended graphene but it is still rather high exceeding $K$ of Si ($K$=145 W/mK) and copper ($K$=400 W/mK). Solving BTE, the authors determined $K$ of free graphene to be ~3000 W/mK near RT. They attributed the reduced experimental value to graphene - substrate coupling and phonon leaking across the interface [78]. An independent study, which used electrical self-heating method, found $K\approx1000$ - 1400 W/mK near RT for graphene nanoribbons (GNRs) for $n$<5 and width 16 nm $< W <$ 52 nm [79]. The breakdown current density of graphene was determined to be ~$10^8$ A/cm$^2$, close to that of CNTs. The authors [79] did not measure the thermal interface resistance of graphene-substrate but rather assumed that it is the same as that of SW-CNT [79]. Table I provides representative experimental data for suspended and supported graphene.

## VII.    Thermal Transport in Few-Layer Graphene

It is interesting to examine evolution of thermal properties of FLG with increasing thickness $H$ (number of atomic planes $n$). One has to clearly distinguish two cases: thermal transport limited by (i) intrinsic properties of FLG lattice, e.g. crystal anharmonicity, and (ii) extrinsic effects, e.g. by phonon – boundary or defect scattering. The optothermal Raman study [74] found that $K$ of suspended uncapped FLG decreases with increasing $n$ approaching the bulk graphite limit (Figure 3a). This evolution of $K$ was explained by considering the intrinsic quasi-2D crystal properties described by the phonon Umklapp scattering [74]. As $n$ in FLG increases – the phonon dispersion changes and more phase-space states become available for phonon scattering leading to $K$ decrease. The phonon scattering from the top and bottom boundaries in suspended FLG is limited if constant $n$ is maintained over the layer length. The small thickness of FLG ($n$<4) also means that phonons do not have transverse component in their group velocity ($\upsilon_\perp = 0$) leading to even weaker $1/\tau_B$ term for phonon scattering from the top and bottom boundaries. In FLG films with $n$>4 the boundary scattering can increase because $\upsilon_\perp \neq 0$ and it is harder to maintain constant $n$ through the whole area of FLG flake resulting in $K$ below the graphite limit. The graphite value recovers for thicker films. One should note that experimental data-points in Figure 3a are normalized to the same width $W$=5 μm. It is not possible to obtain a set of high-quality damage-free FLG samples with varying $n$ and identical $W$ and shape. The normalization procedure was described in details in ref. (74).

Experimentally observed evolution of heat conduction in FLG with $n$ varying from 1 to $n$~4 [74] is in qualitative agreement with the theory for the crystal lattices described by the Fermi-Pasta-Ulam Hamiltonians [56]. Recent non-equilibrium MD calculations for graphene nano-ribbons with the number of planes $n$ from 1 to 8 [80] gave the thickness dependence $K(n)$ in excellent agreement with the experiment [74]. As seen in Figure 3b, $K$ saturates near bulk graphite's value at $n$~4-7. The authors did not observed $K$ dependence on the nano-ribbon width $W$ because $W << \Lambda$ and perfectly periodic boundary conditions were assumed for the edges (i.e. $p$=1). Strong quenching of $K$ as $n$ changes from 1 to 2 is in line with the earlier theoretical predictions [66]. It is also consistent with the experimental dependence of $K$ on the outer-diameter (i.e. number of atomic layers) in MW-CNTs [65, 71]. Another group solved BTE under assumptions that in-plane interactions are described by Tersoff potential while Lennard-Jones potential models interactions between atoms belonging to different layers [81]. They obtained a strong $K$ decrease as $n$ changed from 1 to 2 and slower decrease for $n$>2.

The situation is entirely different for encased graphene where thermal transport is limited by the acoustic phonon scattering from the top and bottom boundaries and disorder, which is unavoidable when FLG is embedded between two layers of dielectrics. A study [82], conducted with the 3-omega technique, found $K \approx 160$ W/mK for encased SLG at $T$=310 K. It increases to ~1000 W/mK for graphite films with the thickness $H$~ 8 nm (Figure 3c). It was also found that for a given $H$, the suppression of $K$ in encased graphene, as compared to bulk graphite, was stronger at low temperature ($T$<150 K) where $K \sim T^{\beta}$, 1.5<$\beta$<2 [82]. Thermal conduction in encased FLG was limited by the rough boundary scattering and disorder penetration through graphene. The presence of the evaporated oxide on top of graphene is known to cause defects in the graphene layer. Correspondingly, $K$ dependence on thickness $H$ was similar to other material system where $K$ is extrinsically limited and scales with $H$. In conventional crystalline thin films, where the thickness $H < \Lambda$, but still much larger than the lattice constant, $K$ grows with $H$ as $K \sim C\upsilon H$ until it reaches the bulk limit $K \sim C\upsilon \Lambda$. A similar scaling with $H$ was observed for encased FLG (Figure 3c) and ultra-thin DLC films (Figure 3d). The overall values of *K* in encased DLC films are much smaller than those for encased FLG, as expected for more disordered materials, but the *K(H)* trend is essentially the same. In ultra-thin DLC, the interface layers are known to be mostly disordered sp$^2$ phase [42]. In both encased FLG and ultra-thin DLC films, the scaling cannot follow exactly the same dependence as in crystalline films because of the influence of disorder and changes of the material properties with *H*.

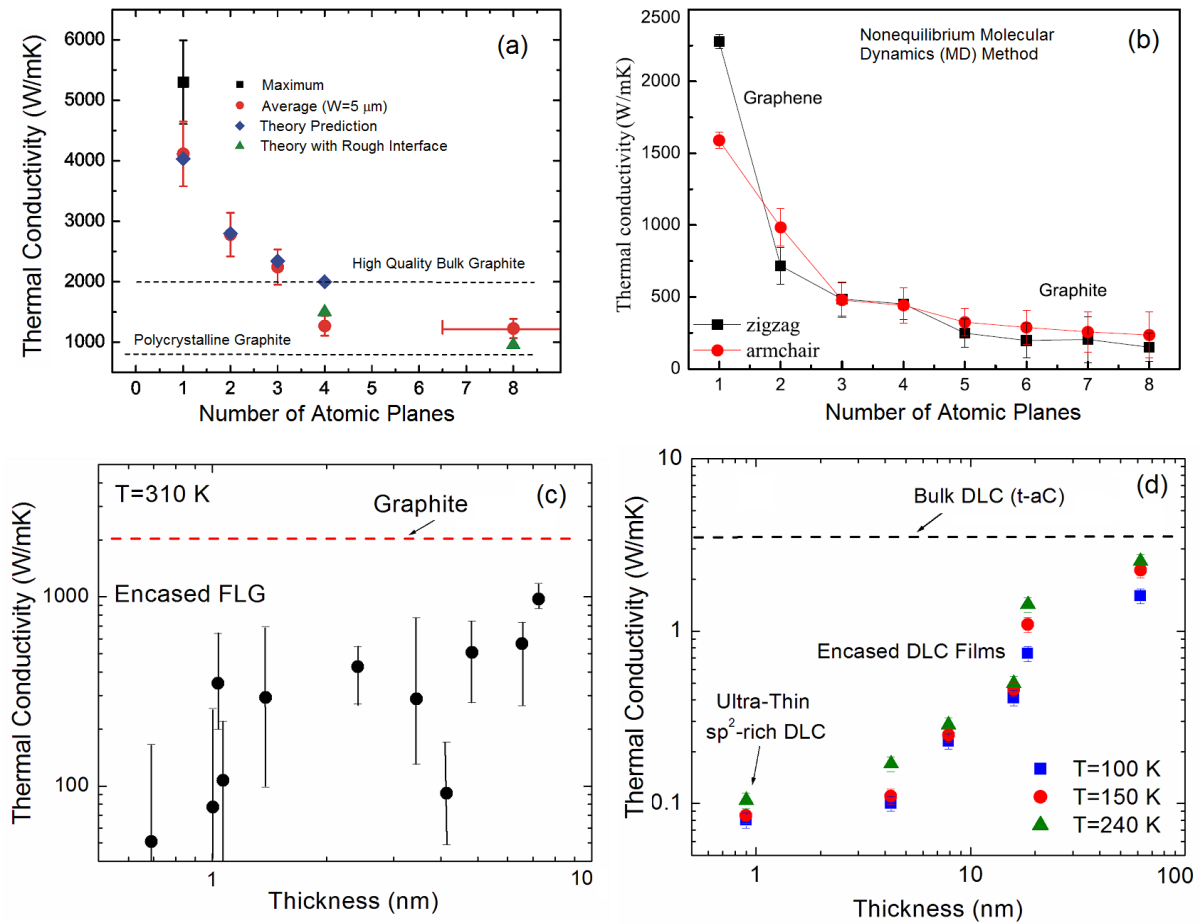

**Figure 3:** Thermal conductivity of quasi-2D carbon materials: intrinsic vs. extrinsic effects. (a) Measured and calculated thermal conductivity of suspended FLG as a function of the number of atomic planes *n*. For *n>4*, *K* can drop below the bulk graphite limit due to the onset of the phonon – boundary scattering from the top and bottom interfaces; *K* recovers for sufficiently thick films. (b) Thermal conductivity of graphene nano-ribbons obtained from MD simulations as a function of *n* showing a similar trend. (c) Measured *K* of encased FLG as a function of the thickness *H*. The transport is dominated by the phonon - boundary scattering and disorder resulting in characteristic *K* scaling with *H*. (d) Thermal conductivity of encased ultra-thin DLC films as a function of the thickness *H*, indicating a similar trend to the encased FLG. Figures (a-d) are adapted from refs. (74), (80), (82) and (42), respectively.

### VIII.    Theory of Heat Conduction in Graphene and CNTs

Measurements of thermal properties of graphene stimulated a surge of interest to theoretical studies of heat conduction in graphene and graphene nanoribbons [83 - 91]. The high-quality suspended FLG also made possible experimental testing of theoretical results obtained for heat conduction in 2D lattices [56-60]. Theoretical description of thermal properties of 2D graphene is closely related to that of CNTs [69]. While analyzing theoretical results one needs to take into account differences between the ballistic ($L<\Lambda$) and diffusive ($L>\Lambda$) transport regimes, and specifics of the intrinsic thermal conductivity in 2D systems (see Box II) related to $K$ divergence with the system size $L$.

Thermal conductivity of *graphene* was addressed, for the first time, within the framework of RTA [23, 61]. It was shown that intrinsic $K$ of graphene should exceed that of bulk graphite when the lateral size of graphene layer becomes sufficiently large (see details in Appendix II). The Klemens theory predicted that the Umklapp-limited $K$ in graphene has logarithmic divergence with the layer or grain size $L$. Substrate coupling reduced $K$ via increase phonon leakage to the substrate and phonon scattering [23, 61]. The analytical expression for graphene was obtained as a special case of the bulk graphite theory [33] with the principle difference being the fact that in graphene the contribution of the low-energy phonons is not limited by cut-off phonon frequency and extends all the way to zero frequency [23, 61]. The theory gave excellent agreement with experimental data for graphite under the assumption that heat is carried mostly by LA and TA phonons, and contributions of ZA phonons are negligible due to their small group velocity $\upsilon \to 0$ in the BZ center and large Gruneisen parameter $\gamma$ [23, 33, 61]. The modified theory with $\gamma$, determined independently for LA and TA modes, provided excellent agreement with experimental data for graphene (Box II). Addition of realistic concentrations of defects in the framework of RTA, allows one to remove the logarithmic divergence of $K$ and obtain meaningful results for graphene [83].

The first equilibrium and non-equilibrium MD simulations determined $K\approx6600$ W/mK for (10, 10) CNT and even a higher $K\approx9000$ W/mK for graphene near RT [66]. It was noted that once graphene layers are stacked in graphite, the interlayer interactions quench $K$ of the system by *an order of magnitude* [66]. In the last few years, a number of MD studies, with Tersoff and Brenner potentials, addressed heat conduction in graphene nanoribbons (GNR) with various length, edge roughness and defect concentration [84-91]. A recent MD study found $K \approx 8000 - 10000$ W/mK at RT for square graphene sheet, which was size independent for $L>5$ nm [84]. For the ribbons with fixed $L=10$ nm and width $W$ varying from 1 to 10 nm, $K$ increased from ~1000 W/mK to 7000 W/mK. Thermal conductivity in GNR with rough edges can be suppressed by orders of magnitude as compared to that in GNR with perfect edges [84, 87]. Table I summarizes $K$ calculated for graphene using different approaches.

An interesting question, which has practical implications, is what carbon low-dimensional material – CNT or graphene – has higher intrinsic $K$. A recent theoretical study [69] has found that thermal conductivity of SW-CNTs $K_{CNT}$ is always below of that of graphene $K_G$ for CNT diameters $d_{CNT}>1$ nm (Figure 4a). The calculation included contributions from all phonon modes – TA, LA and ZA. $K$ of SW-CNTs was found to be ~ $0.8 \times K_G$ for CNTs with the diameter $d_{CNT} \sim 1.5$ nm and gradually increased with $d_{CNT}$ approaching $K_G$ for $d_{CNT} \approx 8$ nm [69]. The calculated $K$ ($d_{CNT}$) is a non-monotonic function, which gives ~2500 W/mK at RT for $L=3$ μm. The ballistic limit for $K_G$ was found to be as high as 12800 W/mK.

### IX. Theoretical and Experimental Uncertainties

An intriguing open question in the theory of phonon transport in graphene is a relative contribution to heat conduction by LA, TA and ZA phonon polarization braches (Box II). There have been opposite view expressed to the importance of ZA phonons, from negligible [23, 33, 61], to dominant [69, 78, 81, 85]. The argument against ZA contributions originates from Klemens' theory, which states that ZA modes have large $\gamma$ [22-23, 61], which defines the scattering strength, and zero group velocity near zone center resulting in negligible contribution to heat transport [23, 33, 61]. The argument for the strong contributions of ZA modes is made on the basis of a selection rule in ideal graphene, which restricts the phase space for phonon-phonon scattering, and increases ZA modes lifetime [85]. However, graphene placement on any substrates and presence of nanoscale corrugations in graphene lattice can break the symmetry selection rule, which restricts ZA phonon scattering. From another side, it is possible that ZA dispersion undergoes modification, e.g. linearization, due to the substrate coupling. An answer to the question of relative contributions may take time, considering that after almost a century of investigations there are still debates about contributions of LA and TA phonons in conventional semiconductors. The measurements of $T^\beta$ dependence alone cannot present an evidence in favor of one or the other phonon contribution because $K(T)$ dependence in graphite is known to depend strongly on the material quality [29-30, 82].

In order to compare directly independent measurements of thermal conductivity of graphene, I reproduce here the measured $K$ as the function of temperature $T$ from ref. (92), and add experimental [16, 17, 76, 93] and theoretical [62, 83] data from other works (See Figure 4b). In this plot, $K$ is larger for graphene than graphite. At high $T$>500 K the difference becomes less pronounced, which is expected as the higher phonon energy levels become populated.

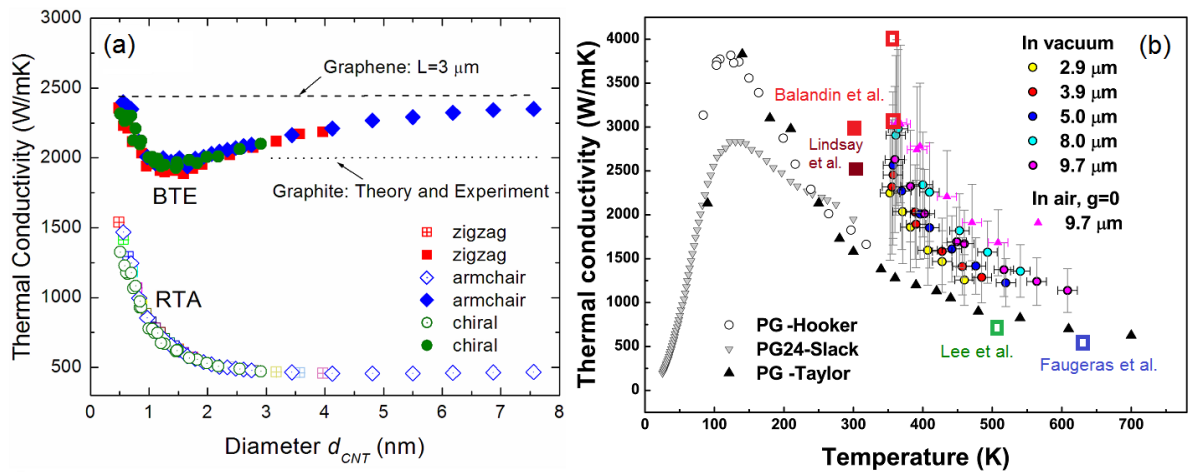

**Figure 4**: Thermal properties of low-dimensional carbon materials. (a) Calculated thermal conductivity of CNTs and graphene indicating that the intrinsic $K$ of CNTs is always lower than that of graphene. (b) Experimental thermal conductivity of graphene as a function of temperature reproduced from ref. (92). Experimental data points from other works are indicated by empty rectangular boxes – red (from refs. (16, 17, 74)), green (ref. (93)), and blue (ref. (76)). The experimental uncertainty of these data points should be comparable to those indicated for ref. (92). The filled red and brown boxes are theoretical data points from refs. (62) and ref. (69), respectively. These two points are for different graphene flake sizes – 3 μm and 5 μm, respectively. Setting $L$=3 μm in ref. (62) would give $K$~2500 W/mK as in ref. (69). Figures (a) and (b) are adapted from refs. (69) and (92), respectively.

Note that some studies have not measured the absorption under the conditions of the experiment [76, 93]. It was recently found that optical absorption in graphene is a strong function of wavelength $\lambda$ owing to the many-body effects [94], which can lead to different absorption in optothermal experiments (see Appendix I). The absorption of 2.3% is observed in the near-IR at ~1 eV. Absorption steadily increases for energies higher than 1.5 eV. The 514.5-nm and 488-nm Raman laser lines correspond to 2.41 eV and 2.54 eV, respectively. The assumption of 2.3% in the Raman measurements with $\lambda > 1.5$ eV leads to underestimated thermal conductivity. To date, the highest values of thermal conductivity of graphene were measured with the Raman optothermal technique. It is difficult to compare directly its accuracy with that of the thermal-bridge or 3-ω techniques [21, 39-42] when the latter are applied to graphene. The Raman technique used for graphene [16-17, 74-77] has a benefit of relative ease of sample preparation and reduced sample contamination. However, its temperature resolution is substantially inferior to the 20-50 mK sensitivity, which can be achieved with the resistance temperature detectors. The Raman data for graphene is often reported with up to 40% uncertainty in the absolute $K$ value. At the same time, the assembly of suspended graphene between two suspended micro-thermometers makes such measurements extremely challenging and results in ambiguity related to the influence of the residual polymeric layers and other defects created during fabrication. More work has to be done with the suspended micro-thermometers to have accurate assessment of the systematic errors in the various techniques (see *Prospective*).

One should keep in mind that comparison of $K$ of graphene and graphite contains ambiguity related to a definition of the graphene thickness $h$. Most studies used $h$=0.34 nm defined by the carbon-bond length. However, this definition is not unique [95-96]. One can introduce $h$ from the inter-atomic potential [95] or start from Young's modulus and tensile strength [96] obtaining $h$ in the range from 0.06 to 0.69 nm, which can shift $K$ of graphene up and down as compared to bulk graphite value [83]. This means that consistent use of $h$=0.34 nm allows for comparison of the results obtained for graphene in different groups. However, it leaves ambiguity while comparing $K$ for graphene and graphite. While the theoretical evidence and results of the optothermal Raman measurements suggest that $K$ of graphene can exceed that of graphite a particular choice of $h$ can shift the $K$ curves up or down in Figures 3 and 4.

## X. Thermal Transport at Graphene-Substrate Interfaces

Thermal boundary resistance (TBR) at the interface of graphene with other materials is a subject of both fundamental science and practical interest. The knowledge of TBR can help in understanding graphene thermal coupling to matrix materials. Controlling TBR is important for graphene's electronic and thermal management applications. TBR is defined as $R_B = [\bar{q}/\Delta T]^{-1}$, where $\Delta T$ is the temperature difference between two sides of the interface. It has non-zero value even at the perfect interfaces due to differences in the phonon DOS - an effect known as Kapitza thermal resistance [97]. The actual TBR is usually higher than Kapitza resistance due to interface imperfections. Heat conduction across graphene or FLG was measured by several different techniques, including electrical 3-ω [98], Raman-electrical [99-100], and optical pump-and-probe [101] methods. The thermo-reflectance technique was used to study graphite interface with Cr, Al, Ti and Au [102]. The RT TBR of ~$10^{-8}$ (Km$^2$/W) was found in most of cases. The studies are in agreement that the cross-plane conductance $G$ or TBR do not reveal a strong dependence on the thickness of FGL or nature of the dielectric or metal substrate. TBR decreases with $T$ following a typical trend for Kapitza resistance [97].

A first-principle calculation of heat transfer between graphene and SiO$_2$, which treated the graphene-substrate coupling as a week van-der-Waals-type interaction, determined the heat transfer coefficient of ~ $2.5 \times 10^7$ (W/m$^2$K) [103]. This translates to TBR of ~$4 \times 10^{-8}$ (Km$^2$/W), which is close to experimental data. Despite agreement on the average TBR, most studies note a significant sample to sample variation in TBR at graphene/SiO$_2$ interface (e.g factor of ~4 for FLG with $n$=5 in ref. (100)). This means that graphene thermal coupling to other materials can depend strongly on the surface roughness, presence or absence of suspended regions in graphene layers, and methods of graphene preparation. MD simulations found that TBR at graphene-oil interface [104, 107] is similar or smaller than that at the graphene-solid interfaces [104 - 107]. Low TBR of graphene with many materials is good news for graphene applications in thermal interface materials (TIMs).

## XI. Graphene and Carbon Based Composites

The needs for improved TIMs in modern electronics and optoelectronics stimulated interest to carbon materials as fillers for TIMs [105-113]. Current TIMs are based on polymers or greases filled with thermally conductive particles such as silver, which require high volume fractions of filler (up to 70%) to achieve $K$ of ~1-5 W/mK of the composite. Carbon materials, which were studied as fillers include CNTs, graphite nanoplatelets (GNPs), graphene oxide nanoparticles (GONs), and graphene flakes derived by chemical processes.

The thermal conductivity enhancement factor $\eta = (K_e - K_b)/K_b$, where $K_e$ is thermal conductivity of the composite material and $K_b$ is thermal conductivity of the initial base material, for composites is shown in Table II.

**Table II: Thermal Conductivity Enhancement in Nano-Carbon Composites**

| Filler | Enhancement | Volume Fraction | Base Material | Refs |
|--------|-------------|-----------------|---------------|------|
| MW-CNT | 160% | 1.0 vol% | oil | 105 |
| SW-CNT | 125% | 1.0 wt% | epoxy | 106 |
| SW-CNT | 200% | 5.0 wt% | epoxy | 108 |
| GNP | 3000% | 25.0 vol% | epoxy | 109 |
| GON | 30% - 80% | 5.0 vol% | glycol; paraffin | 110 |
| Graphene Oxide | 400% | 5.0 wt% | epoxy resin | 111 |
| Graphene | 500% | 5.0 vol% | silver epoxy | 112 |
| Graphene | 1000% | 5.0 vol% | epoxy | 112 |

Despite variations in $\eta$, explained by different base materials and preparation methods, the conclusion is that graphene, CNTs and GON are promising as fillers in terms of the resulting $K_e$. The enhancements are above 100% for 1 wt% of the CNT or graphene loading. This is not achievable with conventional fillers. Graphene demonstrated the highest $\eta$ owing to its geometry and better coupling to base materials [107, 112]. Future TIM applications of carbon materials would depend on many factors including the composite viscosity, coefficient of thermal expansion (CTE), TBR and cost. For epoxy - graphene composites, CTE was found to vary in the range $\sim(5\text{-}30)\times10^{-5}$ (1/$^o$C) and decreased with increasing graphene fraction [111]. Carbon nanoparticles strongly enhance thermal diffusivity of epoxy to $D_T\sim60$ mm$^2$/s at 70 vol% [114]. An important characteristic for TIM applications of graphene is its high temperature stability, which was verified up to 2600 K [115]. The use of liquid-phase exfoliated graphene [116] in advanced TIMs can become the first industry application, which would require large quantities of this material [117].

## XII. Thermoelectric Effects in Graphene

Experimental studied [118] demonstrated that graphene, with the electron mobility ranging from 1000 to 7000 cm$^2$/Vs, have the peak value of the thermoelectric (TE) power (TEP) of $\sim80$ μV/K at RT. The TEP sign, which defines the majority charge carriers, changed from positive to negative as the gate bias crossed the charge neutrality point. Similar results with TEP of ~50-100 µV/K were obtained in other experiments [119-120]. The theory gave consistent results [121]. It was established theoretically that TEP behaves as $1/(n_0)^{1/2}$ at high carrier density $n_0$, but saturates at low densities. TEP scales with the normalized temperature $T/T_F$ and does not depend on the impurity densities ($T_F$ is Fermi temperature) [121]. Calculations [122] reproduced experimental results [100] for Seebeck coefficient $S$ ranging from ~10 to ~100 µV/K for $T$ from ~100 to ~300 K. The theoretical studies of the phonon-drag effects on TEP in bilayer graphene revealed higher $S$ at low $T$ [123].

The efficiency of TE energy conversion is determined by the figure-of-merit $ZT=S^2\sigma T/(K_e+K_p)$. Moderate values of $S$ mean that $ZT$ can only be made practically relevant if $K$ is suppressed. Although graphene reveals extremely high intrinsic $K$, its dominant $K_p$ component can be efficiently suppressed by utilizing graphene ribbons with rough edges or introducing disorder [87, 124-125]. Theoretical studies suggested that $ZT$ can be made as high as ~4 at RT in zigzag GNRs [126]. For comparison, ZT in state-of-art thermoelectrics is ~1 at RT. The improvement in GNR ZT results from strong suppression of $K_p$ due to phonon-edge disorder scattering without substantial deterioration of electron transport [126]. Graphene, with intentionally introduced lattice disorder, e.g. via electron beam irradiation [127], or charged impurities [128] can become an option for TE energy conversion. Graphene reveals interesting TE effects, it has high $S$ compared to elemental semiconductors and the $S$ sign can be changed by the gate bias, instead of doping. However, a possibility of graphene's TE applications is still a subject of debates.

## XIII. Prospective

Carbon materials reveal a unique range of thermal properties with $K$ varying from 0.01 W/mK to above 3000 W/mK near RT. If needed, e.g. for TE applications, $K$ of graphene can be tuned in a wide range by introduction of disorder or edge roughness. Excellent heat conduction properties of graphene are beneficial for all proposed electronic and photonic applications. The transparent FLG electrodes can perform an additional function of removing heat and improving the efficiency of photovoltaic solar cell via reduction of its temperature under illumination. Similarly, FLG serving as interconnects in 3D electronics can simultaneously act as lateral heat spreaders [129]. The demonstrated $K$ enhancement of composites by addition of small volume fractions of liquid-phase exfoliated graphene is promising for TIM applications. Progress in CVD graphene growth on various substrates [130-132] gives hope that one would have a much better control of thermal properties of supported or encased FLG. FLG lateral heat spreaders deposited on GaN wafers and connected to heat sinks were shown to reduce substantially the temperature rise in the high-power density AlGaN/GaN field-effect transistors [133]. Moreover, even if one has to use thin graphite layers instead of SLG to prevent $K$ degradation, he still benefits from the high in-plane $K \approx 2000$ W/mK of graphite, which is exceptional compared to semiconductors, e.g. $K \approx 150$ W/mK in bulk Si and $K \approx 10$ W/mK in Si nanowires at RT [134].

The explosive growth of the new field of thermal properties of graphene and low-dimensional carbons does not allow one to include all pertinent information in a single review. Unique characteristics, which were not discussed in details, include graphene's negative thermal-expansion coefficient $\alpha = (-4.8 \pm 1.0) \times 10^{-6} (\mathrm{K}^{-1})$ for $T < 300$ K, which switches sign at $T \approx 900$ K for SLG and $T \approx 400$ K for BLG [135]. These properties are rooted in the intricacies of the phonon dispersion in graphene and FLG [74, 83, 135-138]. One can expect that graphene thermal properties can be strain-engineered in a way similar to electronic properties [139]. Another important issue is the effect of the defects, grain size and orientation on $K(T)$ of graphene. Recently, two similar studies [140-141] suggested that $K(T) \sim T^{1.4}$ or $\sim T^{1.5}$ dependences prove that ZA modes are dominant in graphene heat transport. However, it is well known [29] that $K(T)$ dependence is strongly influenced by the material quality (Figure 1b). The $K$ values below bulk graphite and $K(T)$ dependence in refs. (140-141) likely indicate polycrystalline graphene with small and misoriented grains or a high concentration of defects due to processing as suggested in ref. (142). Thermal contact resistance of graphene and CNTs, which can affect the accuracy of measurements and change $K$ values [142-144], also deserves more thorough consideration. It has been shown that $K$ of isotopically pure bulk diamond and Si can be substantially improved as compared to their natural abundance [145-146]. The isotope effects in graphene have been considered only computationally [83, 147] and await experimental investigation. Finally, the rise of graphene [13] renewed interest to other carbon allotropes including their prospects for thermal management [55]. The use of complementary electronic and thermal properties of combinations of low-dimensional carbon materials makes the prospects of carbon or hybrid Si-carbon electronics much more feasible.

***Appendix A: Methods for Measuring Thermal Conductivity of Graphene***

Methods of measuring thermal conductivity $K$ can be divided into two groups: steady-state and transient [20]. In transient methods, thermal gradient is recorded as a function of time, enabling fast measurements of the thermal diffusivity $D_T$ over large $T$ ranges. The specific heat $C_p$ and mass density $\rho_m$ have to be determined independently to calculate $K = D_T C_p \rho_m$. If $K$ determines how well materials conduct heat, $D_T$ tells how fast materials conduct heat. Although many methods rely on electrical means for supplying heating power and measuring $T$, there are other techniques where the power is provided with light. In many steady-state methods, $T$ is measured by thermocouples. The transient 3-ω technique for thin films [21] uses $T$ dependence of electrical resistivity for $K$ extraction.

The first experimental study of heat conduction in graphene was made possible by developing an *optothermal Raman technique* (Figure A1a). The heating power $\Delta P$ was provided with laser light focused on a suspended graphene layer connected to heat sinks at its ends (e.g. A1b shows FLG with $n$=2 of rectangular shape suspended across a 3-μm-wide trench in Si wafer). Temperature rise $\Delta T$ in response to $\Delta P$ was determined with a micro-Raman spectrometer. The $G$ peak in graphene's Raman spectrum exhibits strong $T$ dependence. Figure B1c presents the temperature shift in BLG. The inset shows that the optical absorption in graphene is a function of the light wavelength due to many-body effects [94]. The calibration of the spectral position of $G$ peak with $T$ was performed by changing the sample $T$ while using very low laser power to avoid local heating [18]. The calibration curve $\omega_G(T)$ allows one to convert a Raman spectrometer into an "optical thermometer". During $K$ measurements the suspended graphene layer is heated by increasing laser power. Local $T$ rise in graphene is determined as $\Delta T = \Delta \omega_G / \xi_G$, where $\xi_G$ is the $T$ coefficient of $G$ peak. The amount of heat dissipated in graphene can be determined either via measuring the integrated Raman intensity of $G$ peak, as in the original experiments [16], or by a detector placed under the graphene layer, as in the follow up experiments [75, 76]. Since optical absorption in graphene depends on the light wavelength [94] and can be affected by strain, defects, contaminations and near-field or multiple reflection effects for graphene flakes suspended over the trenches it is essential to measure absorption under the conditions of the experiment.

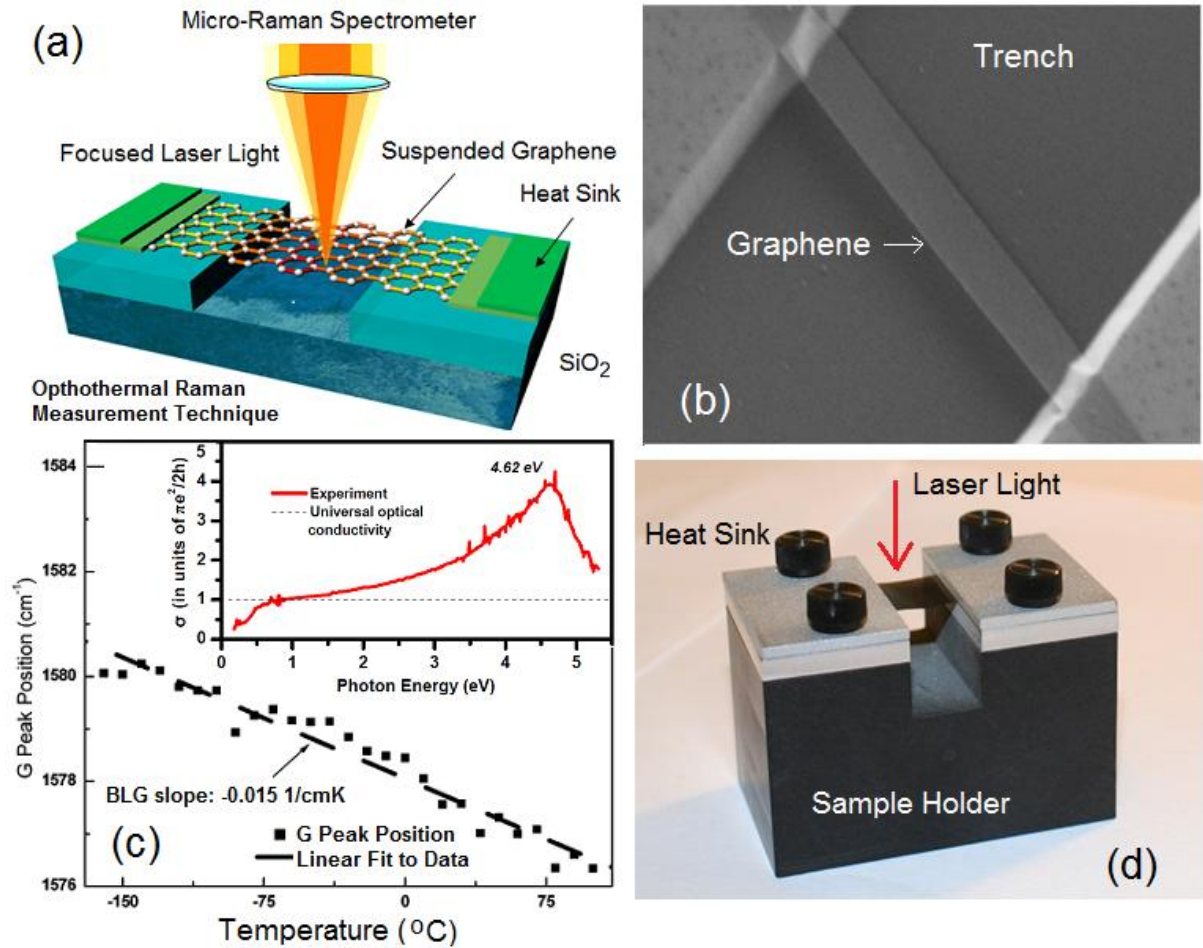

A correlation between $\Delta T$ and $\Delta P$ for graphene samples with a given geometry gives $K$ value via solution of the heat diffusion equation. Large sizes of graphene layers ensure the diffusive transport regime. The suspended portion of graphene is essential for determining $\Delta P$, forming 2D heat front propagating toward the heat sinks, and reducing thermal coupling to the substrate. The method allows one to monitor temperature of Si and $SiO_2$ layer near the trench with suspended graphene from the shift in the position of Si and $SiO_2$ Raman peaks [17]. This can be used to determine the thermal coupling of graphene to $SiO_2$ insulating layer. The optothermal Raman technique for measuring K of graphene is a direct steady-state method. It can be extended to other suspended films (B1d), e.g. graphene films [19], made of materials with pronounced T-dependent Raman signatures. Figure (c) is adapted from ref. (20).

## Appendix II: Unique Features of Heat Conduction in 2D Crystals

Investigation of heat conduction in graphene [16-17] and CNTs [8] raised the issue of ambiguity in definition of the *intrinsic* thermal conductivity for 2D and 1D crystal lattices. It is now accepted that $K$ limited by the crystal anharmonisity alone, referred to as *intrinsic*, has the finite value in 3D bulk crystals [6, 56]. However, the intrinsic $K$ reveals a logarithmic divergence in 2D crystals, $K \sim ln(N)$, and power-law divergence in 1D systems, $K \sim N^\alpha$, with the system size $N$ ($N$ is number of atoms, $0 < \alpha < 1$) [6-7, 56-60]. This anomalous behavior, which leads to infinite $K$ in 1D and 2D systems, is principally different from the *ballistic* heat conduction in structures with the size smaller than the phonon MFP. The logarithmic divergence can be removed by introduction of the *extrinsic* scattering mechanisms such as scattering on defects or by pinning (e.g. coupling to substrates) [56]. Alternatively, one can define the *intrinsic* $K$ of 2D crystals for a given size. There have been indications that for very large sizes of lattices the finite value of intrinsic $K$ can be regained in CNTs or graphene due to the higher-order phonon scattering processes. Nevertheless, the latter has not been conclusively proven yet, and the ambiguity in $K$ is a principally new situation from what we are accustomed to in 3D world.

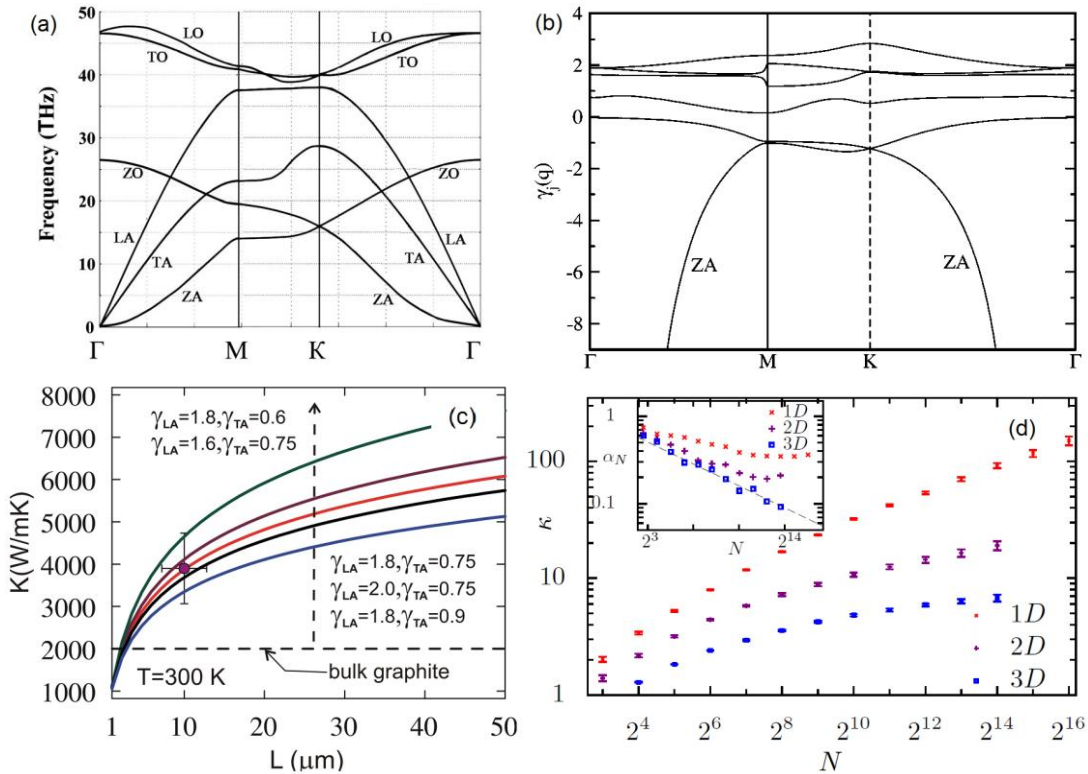

The uniqueness of heat conduction in graphene can be illustrated with an expression, derived by Klemens, for the *intrinsic* Umklapp-limited thermal conductivity of *graphene* [23, 61]:

$$K = (2\pi\gamma^2)^{-1}\rho_m(\upsilon^4/f_mT)\ln(f_m/f_B).$$

Here $\rho_m$ is the mass density, $f_m$ is the upper limit of the phonon frequencies defined by the phonon dispersion and $f_B = \left(M\upsilon^3 f_m/4\pi\gamma^2 k_B TL\right)^{1/2}$ is the size-dependent low-bound cut-off frequency for acoustic phonons, introduced by limiting the phonon MFP with the graphene layer size $L$. Klemens [23, 61] neglected contributions of ZA phonons because of their low group velocity and large Gruneisen parameter $\gamma$. The phonon dispersion and $\gamma$ in graphene are shown in Figures B2a and B2b, respectively. The fundamental $K$ dependence on $L$ obtained from this model is illustrated in Figure B2c. This result is in line with other theoretical approaches [6-7, 56-60], which numerically confirmed that $K$ diverges in 2D anharmonic lattices [56]. Figure B2d shows that anharmonicity is sufficient to have a finite intrinsic $K$ value in 3D crystals (the running slope $\alpha_N$ extracted from *K(N)* dependence goes to zero in 3D case but saturates in 1D and 2D cases). The intrinsic $K$ is defined for an *ideal* graphene without defects. In experiments, $K$ is also limited by extrinsic factors, e.g. point defects, grain boundaries, substrate coupling, etc., and does not grow to the unphysically high values. The figures are adapted from refs. (83), (136), (62), and (56), respectively.

### *Acknowledgements*

I am indebted to K. Saito, L. Lindsay, N. Mingo, C. Dames, R.S. Ruoff, L. Shi, N. Mounet, N. Marzari, B.Q. Ai and T. Heinz for providing figure files. I thank E.P. Pokatilov, D. Nika, C. Dames, L. Shi, D. Cahill, N. Mingo, R. Ruoff, P. Kim, J. Shi, M. Dresselhaus, A. Geim and K. Novoselov for useful discussions. This work was supported by ONR through award N00014-10-1-0224, SRC-DARPA through FCRP Center on Functional Engineered Nano Architectonics (FENA), and DARPA-DMEA under agreement H94003-10-2-1003. Past funding from US AFOSR through contract A9550-08-1-0100 is also acknowledged.